\documentclass[twocolumn,superscriptaddress,aps,prl,showpacs,preprintnumbers,amsmath,amssymb,floatfix]{revtex4-1}

\usepackage[latin9]{inputenc}
\usepackage{color}
\usepackage{dcolumn}
\usepackage{bm}
\usepackage{bbm}
\usepackage{braket}
\usepackage{mathrsfs}
\usepackage{amstext}

\usepackage{txfonts}

\usepackage{amsmath}
\usepackage{dsfont}
\usepackage{euscript}
\usepackage{amssymb}
\usepackage{graphicx}
\usepackage{esint}
\usepackage[unicode=true,
 bookmarks=false,
 breaklinks=false,pdfborder={0 0 1},backref=false,colorlinks=true,citecolor=blue]
 {hyperref}

\makeatletter
\@ifundefined{textcolor}{}
{%
 \definecolor{BLACK}{gray}{0}
 \definecolor{WHITE}{gray}{1}
 \definecolor{RED}{rgb}{1,0,0}
 \definecolor{GREEN}{rgb}{0,1,0}
 \definecolor{BLUE}{rgb}{0,0,1}
 \definecolor{CYAN}{cmyk}{1,0,0,0}
 \definecolor{MAGENTA}{cmyk}{0,1,0,0}
 \definecolor{YELLOW}{cmyk}{0,0,1,0}
}

\newcommand{\eref}[1]{Eq.\,\eqref{#1}}
\newcommand{\fref}[1]{Fig.\,\ref{#1}}

\newcommand\Poly{{\sf{P}}}
\newcommand\PH{{\sf{PH}}}

\newcommand\sharpP{{\sf{\#\Poly}}}

\makeatother

\begin{document}

\title{Exact sampling hardness of Ising spin models}

\author{B.\,Fefferman}
\affiliation{Joint Center for Quantum Information and Computer Science, NIST/University of Maryland, College Park, MD 20742 USA}

\author{M.\,Foss-Feig}
\affiliation{United States Army Research Laboratory, Adelphi, MD 20783, USA}
\affiliation{Joint Quantum Institute, NIST/University of Maryland, College Park, MD 20742 USA}
\affiliation{Joint Center for Quantum Information and Computer Science, NIST/University of Maryland, College Park, MD 20742 USA}

\author{A.\,V.\, Gorshkov}
\affiliation{Joint Quantum Institute, NIST/University of Maryland, College Park, MD 20742 USA}
\affiliation{Joint Center for Quantum Information and Computer Science, NIST/University of Maryland, College Park, MD 20742 USA}

\begin{abstract}

We study the complexity of classically sampling from the output distribution of an Ising spin model, which can be implemented naturally in a variety of atomic, molecular, and optical systems.  In particular, we construct a specific example of an Ising Hamiltonian that---after time evolution starting from a trivial initial state---produces a particular output configuration with probability very nearly proportional to the square of the permanent of a matrix with arbitrary integer entries.  In a similar spirit to BosonSampling, the ability to sample classically from the probability distribution induced by time evolution under this Hamiltonian would imply unlikely complexity theoretic consequences, suggesting that the dynamics of such a spin model cannot be efficiently simulated with a classical computer. Physical Ising spin systems capable of achieving problem-size instances (i.e. qubit numbers) large enough so that classical sampling of the output distribution is classically difficult \emph{in practice} may be achievable in the near future.  Unlike BosonSampling, our current results only imply hardness of \emph{exact} classical sampling, leaving open the important question of whether a much stronger \emph{approximate}-sampling hardness result holds in this context.  As referenced in a recent paper of Bouland, Mancinska, and Zhang \cite{bmz}, our result completes the sampling hardness classification of two-qubit commuting Hamiltonians.

\end{abstract}

\maketitle

\section{Introduction \label{sec: intro}}

It is often taken for granted that quantum computers can efficiently perform certain computational tasks that classical computers cannot.  But finding a quantum task that, on the one hand, admits compelling complexity-theoretic arguments against efficient classical simulation, and on the other hand admits experimental demonstration with technology that is feasible in the near future, remains an important and challenging task in the field of quantum information science.  An extremely exciting line of work, starting with results of Terhal and DiVincenzo and Bremner, Jozsa, and Shepherd, has shown that quantum computers are capable of sampling from distributions that cannot be sampled exactly by randomized classical algorithms \cite{di-terhal,Bremner459}.  The BosonSampling protocol \cite{boson}, proposed by Aaronson and Arkhipov, gives a hardness of sampling result that may be within reach for near-term quantum experiments.  The basic idea is to send photons through a network of linear optical devices, arranged in such a way that the probabilities of typical output configurations of the photons are proportional to the squares of permanents of matrices with independent and Gaussian-distributed random entries.  Given reasonable assumptions about the hardness of computing permanents of such matrices, the ability to efficiently classically sample from any distribution even close (in total variation distance) to this distribution would imply extremely unlikely complexity theoretic consequences.

A number of proof-of-principle experiments implementing BosonSampling have already been carried out \cite{Broome794,Spring798,Tillman_2013,Crespi2013}.  However, a remaining bottleneck to producing an experimentally convincing demonstration of BosonSampling is the technical difficulty of building linear-optical systems that are large enough and clean enough to realize BosonSampling instances for which classical sampling is actually difficult.  By comparison, state preparation and readout of individual spins can be done with high fidelity and relative ease, and the ability to massively parallelize spin-spin interactions between large numbers of qubits is reasonably sophisticated; experiments have successfully implemented some simple instances of the Ising model with system sizes ranging from tens \cite{Richerme_2014} to many hundreds of spins \cite{Bohnet1297}.  Moreover, recent developments in ion-trapping experiments raise the exciting prospect of implementing \emph{arbitrary} Ising interaction graphs in systems of (potentially) many tens of trapped ions \cite{Debnath_2016}. For this reason, finding results analogous to BosonSampling for simple spin models is highly desirable, and potentially affords a simpler route towards the experimental demonstration of an efficient quantum task that, under extremely plausible assumptions about classical complexity theory, cannot be efficiently performed by a classical system \cite{Bremner459,guzik_2015}.

\begin{figure}[!t]
\includegraphics[width=0.99\columnwidth]{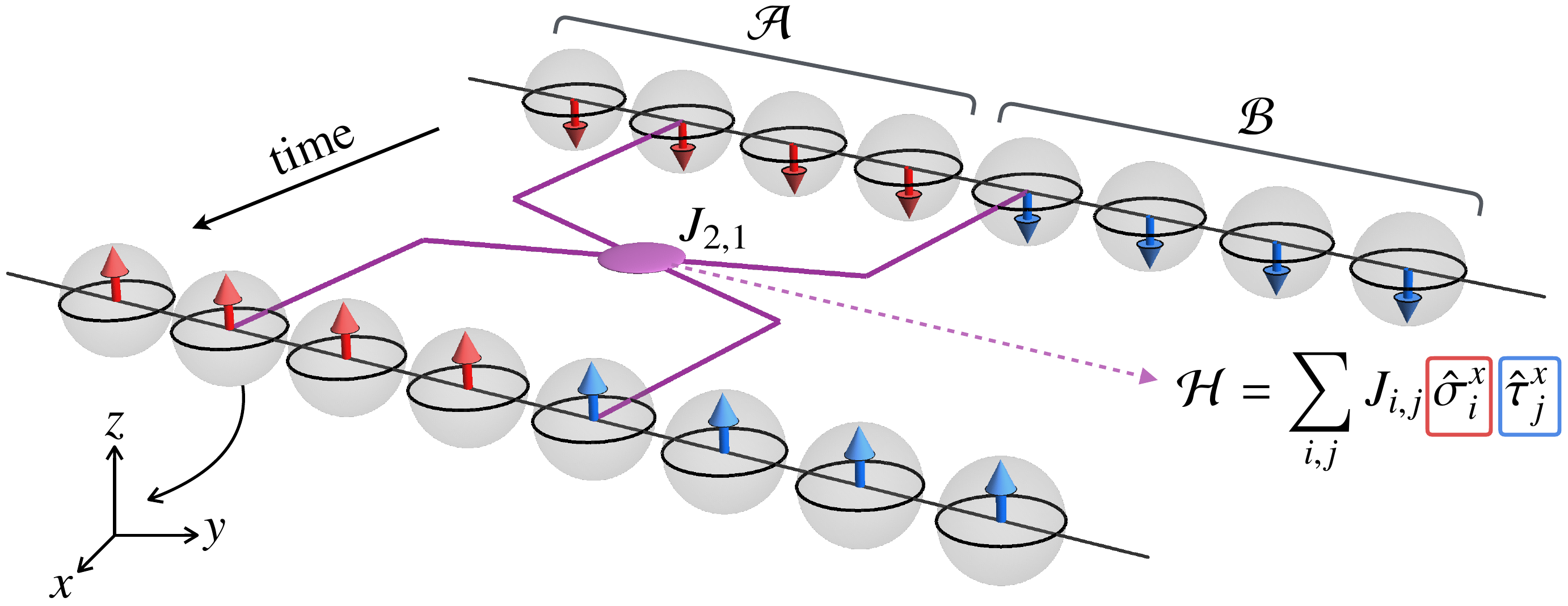}
\centering
\caption{Schematic of the model: (a) Spins in sublattice $\mathcal{A}$ (red) are coupled to spins in sublattice $\mathcal{B}$ (blue) via Ising couplings $\hat{\sigma}^x_i\hat{\tau}^x_j$ and all of them start off in $\ket\downarrow$.  To lowest order in time, the matrix element of the time evolution operator between an initial state with all spins intialized in $\ket\downarrow$ and a final state with all qubits in $\ket\uparrow$ receives contributions in which each spin is flipped precisely once (one such contributing term, between the spin on the second site of $\mathcal{A}$ and the first spin of $\mathcal{B}$, is shown.}
\label{fig:schematic}
\end{figure}

Our goal in this manuscript is to show that the dynamics of an experimentally implementable commuting spin model---the Ising model with no transverse field---can induce an output distribution over the spin states that is hard to sample from classically.  The general strategy, which will be elaborated on below, is to divide a set of Ising spins into two mutually interacting registers, each having $N$ spins (see \fref{fig:schematic}).  The $N$ spins in the first and second register can be placed in correspondence with the $N$ row and column labels, respectively, of an $N\times N$ matrix $J$; each of the $N^2$ pairwise Ising couplings $J_{i,j}$ between a spin ($i$) in one register and a spin ($j$) in the other is a matrix element of $J$.  By initializing the system in a spatially homogeneous product state and then letting it evolve under Ising interactions for a short time, it can be shown that a single probability of the output distribution induced by measurement is proportional to the square of the permanent of $J$, plus an $o(1)$ correction.  This is enough, using a tool known as ``Stockmeyer counting'' \cite{stockmeyer}, to imply a hardness of ``exact sampling'' result: no efficient classical randomized algorithm can sample from \emph{exactly} this distribution, under a ubiquitous hardness assumption (namely, that the Polynomial-time Hierarchy does not collapse).  Note that in a recent paper \cite{guzik_2015}, BosonSampling was directly generalized to the context of spin Hamiltonians.  However, our work encounters the permanent in a fundamentally different way; an important difference is that our results do not rely on a ``diluteness criterion'', and thus $N$ is set by---as opposed to much less than---the number of physical qubits.  Much like other ``exact sampling'' results, our result also demonstrates hardness to classically sample from any distribution in which all probabilities are within a constant multiplicative factor of the ideal quantum distribution.  However, unlike BosonSampling, a recent proposal of Bremner, Montanaro and Shepherd (sometimes called ``IQP'' sampling), and Quantum Fourier Sampling, it is not yet clear whether the distributions we consider can be used to show an ``approximate-sampling'' hardness result \cite{boson, Bremner459, FU15}.  This would show something far stronger: there is no classical algorithm that can sample from any distribution inverse polynomial in total variation distance from the ideal quantum distribution. 

\section{The model}

The model we consider consists of $2N$ spin-1/2 particles, which we divide into two sublattices of $N$ spins each, denoted $\mathcal{A}$ and $\mathcal{B}$ (blue and red spins in Fig.\,\ref{fig:schematic}).  We consider quench dynamics under an Ising Hamiltonian with exclusively two-body \emph{inter-sublattice} interactions (but no interactions within either sublattice), which can take arbitrary integer values,
\begin{align}\label{eq:isinghamiltonian}
\mathcal{H}=\sum_{i,j}J_{i,j}\hat{\sigma}^{x}_i\hat{\tau}_j^x.
\end{align}
Here, Pauli operators $\hat{\sigma}$ act on the spins of sublattice $\mathcal{A}$, while Pauli operators $\hat{\tau}$ act on the spins of sublattice $\mathcal{B}$.  These spins could be, for example, two subsets of ions in a Paul trap, where the $\ket\downarrow$ and $\ket\uparrow$ are, respectively, the electronic ground state and some long-lived metastable state (in general either an excited hyperfine level of the electronic ground-state manifold or a dipole-forbidden optical excitation).  The Ising interactions can then be implemented via a spatially-structured M\o{}lmer-S\o{}rensen interaction \cite{PhysRevLett.82.1835,Korenblit_2012,Debnath_2016}.

We consider a quantum quench in which the system is initialized at time $t=0$ with all of the spins (in both registers) in the spin-down state along the $z$-direction,
\begin{align}
|\psi(0)\rangle=\bigotimes_{i\in\mathcal{A}}\ket{\downarrow}_i\bigotimes_{j\in\mathcal{B}}\ket{\downarrow}_j.
\end{align}
We then allow the system to evolve under the Hamiltonian in \eref{eq:isinghamiltonian} for a time $t$. 
%
%
%
%
%
%

%
%
%

%
%
%
%

\section{Output distribution}\label{sec:distribution}

\begin{figure}[t!]
\includegraphics[width=0.85\columnwidth]{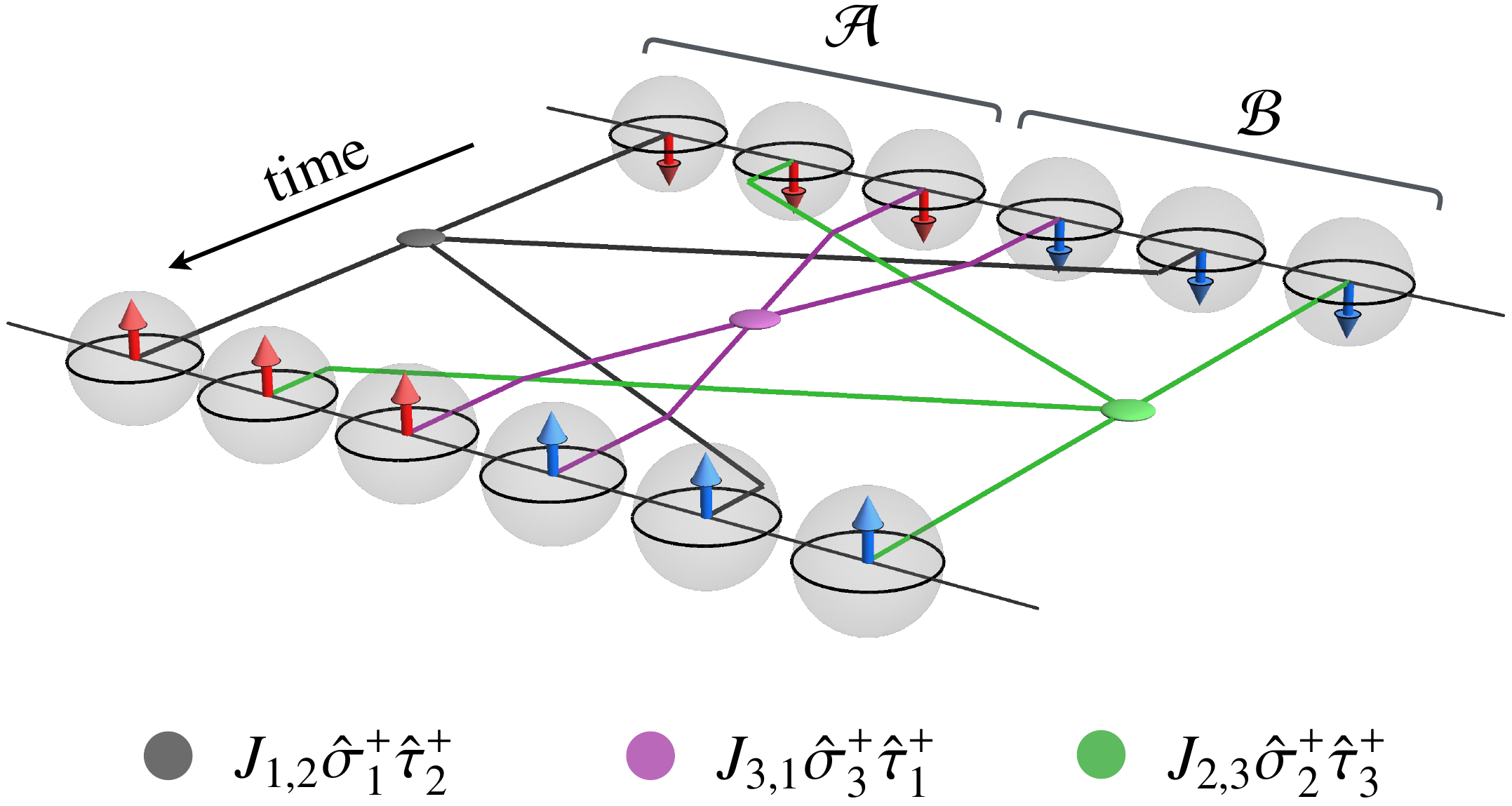}
\centering
\caption{Example of a single term contributing to the matrix element $M_t$ at lowest order in time ($t^N$, here with $N=3$).  Here, all spins are flipped from down to up by a particular pairing off of the spins between the $\mathcal{A}$ and $\mathcal{B}$ sublattices.  The depicted process contributes a term $(J_{1,2}\times J_{3,1}\times J_{2,3})\times (t^3/3!)$ to $M_t$. The set of all possible ways to pair the spins in sublattice $\mathcal{A}$ with the spins in sublattice $\mathcal{B}$ is in one-to-one correspondence with terms in the permanent of the matrix $J_{i,j}$, and thus $M_t$ is proportional to this permanent.}
\label{fig:term}
\end{figure}

After evolution for a time $t$ under the action of $\mathcal{H}$, measurement in the $z$ basis samples from the induced probability distribution
\begin{align}
P_t(\sigma_1&,\dots,\sigma_N,\tau_1,\dots,\tau_N)\nonumber\\
\label{eqn:probabilities}
=&|\bra{\sigma_1,\dots,\sigma_N,\tau_1,\dots,\tau_N }\exp(-i\mathcal{H}t)\ket{\downarrow,\dots,\downarrow} |^2,
\end{align}
where $\sigma_j,\tau_j=\,\downarrow\,,\uparrow$. We are interested in just one such probability,
\begin{align}
P_t\equiv P_t(\uparrow,\dots,\uparrow)=|\bra{\uparrow, \dots, \uparrow }\exp(-it\mathcal{H})\ket{\downarrow,\dots,\downarrow}|^2\equiv|M_t|^2,\nonumber
\end{align}
to end in the state with all spins in both registers pointing up.  By writing an individual term in the Hamiltonian as
\begin{align}
\hat{\sigma}^{x}_i\hat{\tau}_j^x=\hat{\sigma}^{+}_{i}\hat{\tau}_{j}^{+} + \hat{\sigma}^{+}_{i}\hat{\tau}_{j}^{-} + \hat{\sigma}^{-}_{i}\hat{\tau}_{j}^{+} + \hat{\sigma}^{-}_{i}\hat{\tau}_{j}^{-},
\end{align}
it is straightforward to see that repeated applications of $\mathcal{H}$, and thus time evolution, generates population in all possible spin states in the $z$ basis. Expanding $e^{-i\mathcal{H}t}$ as a power series in time, the lowest-order in time non-vanishing contribution to the matrix element $M_t=\bra{\uparrow,\dots,\uparrow}\exp(-it\mathcal{H})\ket{\downarrow,\dots,\downarrow}$ arises at order $t^N$, because every spin needs to be flipped at least once.  The contributing terms contain exactly $N$ powers of operators $\hat{\sigma}^+_i\hat{\tau}_j^+$, with no repetitions of the indices $i$ and $j$, so that each qubit gets flipped from $\ket\downarrow$ to $\ket\uparrow$ exactly one time; see Fig.\ \ref{fig:term} for an illustration of such a term for $N=3$.  It is straightforward to show that, to order $t^{N}$, the matrix element $M_t$ is given by
\begin{align}
M_t&=\frac{(-it)^{N}}{N!}\times N!\sum_{\sigma}\prod_{j=1}^N J_{\sigma(j),j}+O(t^{N+2})\nonumber\\
&=(-it)^{N}{\rm Per}(J) + O(t^{N+2}),
\end{align}
where the summation is over all permutations $\sigma$ of the integers $i=1,\dots,N$.  As a result, and defining $\mathscr{P}=|{\rm Per}(J)|^2$, we have
\begin{align}
P_t=t^{2N}\big(\mathscr{P}+O(t^2)\big).
\end{align}
We next aim to place a constraint on how $t$ must scale with $N$ in order to ensure that the $O(t^2)$ additive error to the permanent is $o(1)$ with respect to the system size $N$.

\section{Higher orders in time\label{sec:bounding}}

As discussed above, the lowest-order in time contribution to the matrix element $M_t$ comes at order $N$.  It is not hard to see that all other contributing terms occur at order $m$ such that $m-N$ is a positive \emph{even} integer.  In particular, take $N_{+-}$ to be the number of times an operator $\hat{\sigma}^+_i\hat{\tau}^-_j$ occurs inside the matrix element, and similarly for $N_{-+}$, $N_{++}$, and $N_{--}$, such that $N_{++}+N_{--}+N_{+-}+N_{-+}=m$.  Since we need to flip the same number of qubits in both registers, we must have $N_{+-}=N_{-+}$.  Also, the total number of flipped qubits is equal to $2(N_{++}-N_{--})$, and since all qubits need to be flipped, we have $N_{++}-N_{--}=N$.  Now, defining $p(n)$ to be the parity of the integer $n$, we have
\begin{align}
p(m)&=p(N_{++}+N_{--}+2N_{+-})\nonumber\\
&=p(N_{++}+N_{--})\nonumber\\
&=p(N_{++}-N_{--})\nonumber\\
&=P(N),
\end{align}
which shows that $m-N$ is an even integer.  The matrix element in question can therefore be expanded as
\begin{align}
M_t=\sum_{\alpha=0}^{\infty}\bra{\uparrow,\dots,\uparrow}\frac{(-it\mathcal{H})^{N+2\alpha}}{(N+2\alpha)!}\ket{\downarrow,\dots,\downarrow}\equiv\sum_{\alpha=0}^{\infty}M_{t}^{(\alpha)},
\end{align}
and from above we have
\begin{align}
M_{t}^{(0)}=(-it)^{N}{\rm Per}(J).
\end{align}
Defining $\delta{M}_t=\sum_{\alpha=1}^{\infty}M_{t}^{(\alpha)}$, such that $M_t=M_{t}^{(0)}+\delta M_t$, we can write
\begin{align}
P_t&=|M_{t}^{(0)}|^2+2\Re[M_{t}^{(0)}\delta M_t]+|\delta M_t|^2\nonumber\\
&=t^{2N}\big(\mathscr{P}+\eta_t\big),
\end{align}
where
\begin{align}
\eta_t&\equiv(2\Re[M_{t}^{(0)}\delta M_t]+|\delta M_t|^2)/t^{2N}\nonumber\\
\label{eq:eta_ineq}
&\leq |\delta M_t|(2|M_{t}^{(0)}|+|\delta M_t|)/t^{2N}.
\end{align}
For notational simplicity, here we will assume that the entries of $J$ are drawn from the set $\{-1,0,1\}$; note that nothing about our argument would change if arbitrary integers were used, except that the time $t$ would be rescaled in the bounds below by $\max(J_{i,j})$. Using $\bra{\uparrow,\dots,\uparrow}\mathcal{H}^{m}\ket{\downarrow,\dots,\downarrow}\leq N^{2m}\lVert\hat{\sigma}^{x}\rVert^{2m}=N^{2m}$, $M_t^{(\alpha)}$ can be bounded as $|M_{t}^{(\alpha)}|\leq (N^2t)^{N+2\alpha}/(N+2\alpha)!$. Therefore,
\begin{align}
\label{eq:ineq1}
|M_t^{(0)}|&\leq\frac{(N^2t)^{N}}{N!},\\
\label{eq:ineq2}
|\delta M_t|&\leq\frac{(N^2 t)^{N}}{N!}\sum_{\alpha=1}^{\infty} (N^4 t^2)^{\alpha}\leq 2 \frac{(N^2 t)^{N}}{N!}(N^4 t^2).
\end{align}
The final inequality in \eref{eq:ineq2} is valid for $t^2\leq1/(2N^4)$, because $0\leq\sum_{\alpha=1}^{\infty}x^{\alpha}\leq 2x$ whenever $0\leq x \leq 1/2$.  Plugging Eqs.\,(\ref{eq:ineq1},\ref{eq:ineq2}) into \eref{eq:eta_ineq} leads to
\begin{align}
\eta_t&\leq 4N^4  t^2 \frac{N^{4N}}{(N!)^2}\bigg(1+N^4  t^2\bigg)\\
&\leq 6N^4  t^2 \frac{N^{4N}}{(N!)^2} \leq t^2{\rm{poly}(N)} e^{2N(\ln N+1)},
\end{align}
with the final inequality obtained by Stirling's approximation.  It follows immediately that $\eta_t=o(1)$ is guaranteed as long as
\begin{align}
t=o(e^{-2N\ln N}).
\end{align}
\section{Hardness of sampling}
Here we prove our main theorem, establishing a very unlikely complexity theoretic consequence which would arise naturally from the presumed existence of a classical algorithm that samples exactly from the output distribution described in the prior sections.  Similar arguments to the one sketched here are implicit in other works on quantum hardness of sampling results starting with the BosonSampling proposal \cite{boson}.

We first begin with a very brief overview of the computational complexity theoretic components necessary to understand this hardness of sampling result.  Computing exactly the permanent of an $N \times N$ matrix $X$ with integer entries is as hard as computing the number of satisfying assignments to a boolean formula.  We therefore say it is a $\sharpP$-hard problem, as established by Valiant \cite{valiant}.  When $X$ has nonnegative integer entries this problem is also in $\sharpP$.

For our purposes, we will be interested in the complexity of computing {\emph{multiplicative estimates}} to the permanent.  We say an algorithm $\mathscr{A}$ efficiently computes a multiplicative estimate to a function $f$ if, given input $x$, the output of $\mathscr{A}$ is within a $1\pm\epsilon$ multiplicative factor of $f(x)$ in time polynomial in $N$ and $1/\epsilon$.  A famous result of Jerrum, Sinclair and Vigoda gives an algorithm for efficiently computing a multiplicative estimate to the permanent of a matrix with {\sl nonnegative} entries \cite{jsv}.  On the other hand, it can be shown using a binary search and padding argument that computing such an estimate to the permanent (or even the square of the permanent) of a matrix with general integer entries is in fact $\sharpP$-hard (see e.g., \cite{boson,scottSharp}).  Therefore computing these estimates are as hard as computing the permanent exactly. How powerful is $\sharpP$?  We know from Toda's Theorem that any problem in the Polynomial-time hierarchy, or $\PH$, can be solved using the ability to solve a $\sharpP$-hard problem \cite{toda}.  Being a bit more formal, Toda's theorem tells us that $\PH\subseteq\Poly^{\sharpP}$. 

Now, for any $N \times N$ matrix $X$ define $\mathcal{D}_X$ to be the outcome distribution from Section \ref{sec:distribution} that arises from starting in the $\ket{\downarrow, \dots,\downarrow}$ state, evolving for a particular time $t$ under the action of the Hamiltonian from \eref{eq:isinghamiltonian} with coupling constants $J_{i,j}$ set to the entries of $X$, and measuring in the $z$ basis.  As shown in Sections \ref{sec:distribution} and \ref{sec:bounding}, the probability of observing the $\ket{\uparrow,\dots,\uparrow}$ outcome at time $t$ is proportional to the square of the permanent of $X$ plus an $o(1)$ correction, provided that $t$ is chosen to be $o(e^{-2N\ln N})$.  Notice that this probability is exponentially small.  Therefore, to get any reasonable estimate by repeated sampling we would need an exponential number of samples.  Indeed, \emph{this does not imply} an efficient quantum algorithm for computing the permanent.  Nonetheless, we can use the fact that a single exponentially small amplitude is proportional to the permanent to argue about the classical intractability of sampling from this distribution.  
  
Suppose we have an efficient classical sampler which samples from the same distribution.  We define this to be an efficient randomized algorithm that takes as input an $N \times N$ integer matrix $X$ and outputs a sample from the distribution $\mathcal{D}_{X}$. A classic result of Stockmeyer gives an algorithm for computing a multiplicative estimate to the probability of any given outcome of an efficient classical sampler in the third level of the $\PH$, or $\sf\Sigma_3$ \cite{stockmeyer}.  Using this result, together with the presumed existence of an efficient classical sampler for our quantum distribution, we can compute a multiplicative estimate to the square of the permanent of an arbitrary integer matrix in the third level of the $\PH$.  As mentioned above, this is a $\sharpP$-hard problem.  This tells us we can solve any problem in $\sharpP$ in the third level of the Polynomial-time hierarchy, or formally, that $\Poly^{\sharpP}\subseteq\sf\Sigma_3$.  Combining this with Toda's theorem, we have that $\PH\subseteq\Poly^{\sharpP}\subseteq\Sigma_3$, and so the entire Polynomial-time Hierarchy collapses to the third level, as claimed.  Therefore, it is very unlikely that an efficient classical sampler for the distribution with probabilities given by \eref{eqn:probabilities} exists.


\section{Discussion and implications}

These results extend several key ideas of BosonSampling to the context of spin dynamics under Ising spin Hamiltonians.  Just like non-interacting bosons, the Ising model without a transverse field is often viewed---from the perspective of many-body quantum physics---to be trivial, since it can be trivially diagonalized.  However, just as with non-interacting bosons, this point of view stems from a restricted notion of what it means to ``simulate'' a quantum system.  As in the case of non-interacting bosons, it is indeed classically efficient to compute low-order correlation functions of operators in the model we study \cite{kastner_2013,PhysRevA.87.042101}, but sampling from the output distribution is simply a more general (and less trivial) task.  

Another interesting motivation for our result comes from the desire to classify all two-qubit commuting Hamiltonians.  Suppose we start in a computational basis state of $n$ qubits, and can apply a fixed two-qubit Hamiltonian to any pair of qubits.  A recent result of Bouland, Mancinska, and Zhang gave a hardness of sampling classification for this model \cite{bmz}.  They prove, in all cases except the one we consider (in which the two qubit Hamiltonian is $X\otimes X$) that the corresponding sampling task is classically hard, as long as the commuting Hamiltonian is capable of generating entanglement from a computational basis state.  Otherwise, the output is in a product state and clearly classically simulable.  Thus our hardness result completes the sampling hardness classification of the complete class of two-qubit commuting Hamiltonians (see their paper for additional details \cite{bmz}).

\section{Acknowledgments}

We thank A. Deshpande and A. Bouland for helpful discussions. We also thank A. Bouland, Laura Mancinska and Xue Zhang for sharing an early version of their results.  A.V.G. acknowledges support by ARL CDQI, ARO MURI, NSF QIS, ARO, NSF PFC at JQI, and AFOSR. This material is based upon work supported by, or in part by, the U. S. Army Research Laboratory and the U. S. Army Research Office under contract/grant number 025989-001.  


%



\end{document}